\documentclass[preprint,aps,amsmath,byrevtex,prd,titlepage,
nofootinbib]{revtex4-1}

\usepackage{dcolumn}
\usepackage{amsmath}
\usepackage{amssymb}
\usepackage{bm}
\usepackage{hyperref}
\usepackage{graphicx}
\usepackage{slashed}

\newcommand{\beq}{\begin{equation}}
\newcommand{\eeq}{\end{equation}}
\newcommand{\eq}[1]{Eq.~(\ref{#1})}

\begin{document}

\title {Light-by-Light Scattering Single-Logarithmic Corrections to Hyperfine Splitting in Muonium}
\author {Michael I. Eides}
\altaffiliation[Also at ]{the Petersburg Nuclear Physics Institute,
Gatchina, St.Petersburg 188300, Russia}
\email[Email address: ]{eides@pa.uky.edu, eides@thd.pnpi.spb.ru}
\affiliation{Department of Physics and Astronomy,
University of Kentucky, Lexington, KY 40506, USA}
\author{Valery A. Shelyuto}
\email[Email address: ]{shelyuto@vniim.ru}
\affiliation{D. I.  Mendeleyev Institute for Metrology,
St.Petersburg 190005, Russia}

\begin{abstract}
We consider three-loop radiative-recoil corrections to hyperfine splitting in muonium generated by the gauge invariant set of diagrams with virtual light-by-light scattering block. These corrections are enhanced by the large logarithms of the electron-muon mass ratio. We present the results of an analytic calculation of the single-logarithmic radiative-recoil  corrections of order $\alpha^2(Z\alpha)(m/M)E_F$ to hyperfine splitting in muonium generated by these diagrams.

\end{abstract}


\preprint{UK/12-11}

\maketitle

\section{Introduction}

\begin{figure}[htb]
\includegraphics
[height=3cm]
{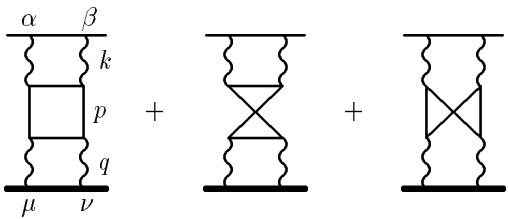}
\caption{\label{lblrec}
Diagrams with light-by-light scattering block}
\end{figure}

Muonium is one of the best studied purely electrodynamic bound states. The hyperfine splitting (HFS) in the ground state of muonium is measured \cite{mbb,lbdd} with error bars in the ballpark of 16-51 Hz. A new higher accuracy measurement of muonium HFS is now planned at J-PARC, Japan \cite{shimomura}. The results of QED calculations of the HFS interval are usually organized in the form of a perturbation theory expansion in $\alpha$, $Z\alpha$, $m_e/m_\mu$. Some of the terms in this expansion are enhanced by large logarithms of the fine structure constant and/or electron-muon mass ratio. The current theoretical uncertainty of the HFS interval is  estimated to be about 70-100 Hz, respective relative error does not exceed $2.3\times10^{-8}$ (see discussions in \cite{egs2001,egs2007,mtn2012}). Still unknown three-loop purely radiative corrections, three-loop radiative-recoil corrections, and nonlogarithmic recoil corrections (see detailed discussion in \cite{egs2007,mtn2012}) are the main sources of the theoretical uncertainty. Measurement of the HFS in muonium is currently the best way to determine the value of the electron-muon mass ratio. The value of $\alpha^2(m_\mu/m_e)$ is obtained from comparison of the HFS theory and experiment with the uncertainty that is dominated by the $2.3\times 10^{-8}$ relative uncertainty of the HFS theory \cite{mtn2012}. Improvement of the HFS theory would allow further reduction of the uncertainty of the electron-muon mass ratio. A detailed analysis \cite{egs2001,egs2007} shows that reduction of the theoretical error of HFS theory in muonium to about $10$ Hz is a  realistic goal. As a step in this direction we consider below three-loop radiative-recoil contributions to HFS generated by the light-by-light (LBL) scattering diagrams in Fig. \ref{lblrec} (and by three more diagrams with the crossed photon lines). The radiative-recoil corrections due to the LBL diagrams in Fig. \ref{lblrec} are additionally enhanced by the large logarithm of the electron-muon mass ratio. The logarithm squared contribution was calculated long time ago \cite{eks89}. Below we calculate the single-logarithmic radiative-recoil contribution.

We will follow the general approach to calculation of three-loop radiative-recoil corrections to HFS developed in \cite{es0,eks89,egs01,egs03,egs04,esprd2009,esprl2009,esjetp2010} and start with the general expression for the LBL scattering contribution  in Fig. \ref{lblrec} (see, e.g., \cite{egs2001,egs2007})

\beq \label{lblskel}
\begin{split}
\Delta E=&\frac{\alpha^2(Z\alpha)}{\pi^3}E_F\frac{m}{M}\left(-\frac{3M^2}{256}\right)\int \frac{d^4 k}{i\pi^2k^4}
\left(\frac{1}{k^2+2mk_0}+\frac{1}{k^2-2mk_0}\right)
\langle\gamma^\alpha\slashed k \gamma^\beta\rangle
\\
\times&\int \frac{d^4q}{i\pi^2 q^4}
\left(\frac{1}{q^2+2Mq_0}+\frac{1}{q^2-2Mq_0}\right)
\langle\gamma^\mu\slashed q \gamma^\nu\rangle S_{\alpha\beta\mu\nu},
\end{split}
\eeq

\noindent
where $k^\mu$ is the four-momentum carried by the upper photon lines, $q^\mu$ is the four-momentum carried by the lower photon lines, $m$ is the electron mass, $M$ is the muon mass, $Z=1$ is the muon charge in terms of the electron charged used for classification of different contributions, and  $S_{\alpha\beta\mu\nu}$ is the light-by-light scattering tensor. The Fermi energy is defined as

\beq
E_F=\frac{8}{3}(Z\alpha)^4 \frac{m}{M}\left(\frac{m_r}{m}\right)^3mc^2,
\eeq

\noindent
where $m_r$ is the reduced mass. The angle brackets in \eq{lblskel} denote the projection of the $\gamma$-matrix structures on the HFS interval (difference between the states with the total spin one and zero).

The contributions to HFS of the first two diagrams coincide and with account for three more diagrams with crossed photon lines not shown explicitly in Fig. \ref{lblrec} we can represent the LBL block as a sum of two contributions, the first one corresponding to the first two (ladder) diagrams in  Fig. \ref{lblrec} and the second one corresponding to the last (crossed) diagram in Fig. \ref{lblrec}

\beq \label{lblintegral}
S_{\alpha\beta\mu\nu}=\int \frac{d^4p}{i\pi^2}\left(2L_{\alpha\beta\mu\nu}+C_{\alpha\beta\mu\nu}\right),
\eeq

\noindent
where

\beq \label{laddtr}
L_{\alpha \beta \mu \nu} = Tr\left(\gamma_{\mu}\frac{1}{{\slashed p}-{\slashed q}-m}\gamma_{\nu}\frac{1}{{\slashed p}-m}\gamma_{\beta} \frac{1}{{\slashed p}-{\slashed k}-m}\gamma_{\alpha}\frac{1}{{\slashed p}-m}\right),
\eeq
\beq \label{crstr}
C_{\alpha \beta \mu \nu} = Tr\left(\gamma_{\mu}\frac{1}{{\slashed p}-{\slashed q}-m}\gamma_{\beta}\frac{1}{{\slashed p}-{\slashed q}-{\slashed k}-m}\gamma_{\nu}\frac{1}{{\slashed p}-{\slashed k}-m}\gamma_{\alpha} \frac{1}{{\slashed p}-m}\right).
\eeq

The integral in \eq{lblskel} contains both nonrecoil and recoil corrections to HFS that are partially already calculated (see \cite{egs2001,egs2007} for a collection of these results)

\beq \label{structure}
\Delta E=\frac{\alpha^2(Z\alpha)}{\pi}E_F[-0.472~514~(1)]
+\frac{\alpha^2(Z\alpha)}{\pi^3}E_F\frac{m}{M}
\left(\frac{9}{4}\ln^2{\frac{{M}}{m}}
+ C_1\ln{\frac{{M}}{m}} +C_0\right).
\eeq

The leading nonrecoil term in \eq{structure} is generated by the nonrelativistic pole in the muon propagator

\beq  \label{NR-pole}
\frac{1}{q^2+2Mq_0+i0} \longrightarrow  -\frac{i\pi}{M}\delta (q_0),
\eeq

\noindent
and was calculated in \cite{eks1991,kn199496}. This is a numerically dominant contribution and it should be extracted analytically from the expression in \eq{lblskel} before calculation of the radiative-recoil corrections.

The diagrams in Fig. \ref{lblrec} contain three loop integrations and each of them could in principle generate a large logarithm of the electron-muon mass ratio. The strongly ordered region of integration momenta $m\ll k\ll p\ll q\ll M$  would produce logarithm cubed contribution but it turns into zero due to the tensor structure of the LBL block and fermion factors in this region \cite{es0}, see below. The large logarithm squared, calculated in \cite{eks89} arises from two integration regions, $m\ll k\sim p\ll q\ll M$ and $m\ll k\ll p\sim q\ll M$. Our current goal is to calculate single-logarithmic contribution is \eq{structure}, and as a preliminary step we would like to separate the large logarithm squared contribution. First we recalculate the logarithm cubed and  logarithm squared corrections in an intuitively transparent way. We will elucidate the origin of different contributions to the logarithm squared term what will help us to derive a convenient expression for calculation of the single-logarithmic contributions.

\section{Leading Logarithmic Contribution}

We start calculation of the integral in \eq{lblskel} with the loop integration in the LBL scattering block. This block is a four-index tensor that is a function of four-momenta $k$ and $q$. Tensor  indices are contracted with the conserved antisymmetric electron and muon factors

\beq \label{skobki}
\begin{split}
&\langle\gamma^{\alpha}{\slashed k}\gamma^{\beta}\rangle
=-\langle\gamma^{\beta}{\slashed k}\gamma^{\alpha}\rangle,
\qquad
k_{\alpha}\langle\gamma^{\alpha}{\slashed k}\gamma^{\beta}\rangle=
k_{\beta}\langle\gamma^{\alpha}{\slashed k}\gamma^{\beta}\rangle=0,
\\
&\langle\gamma^{\mu}{\slashed q}\gamma^{\nu}\rangle
=-\langle\gamma^{\nu}{\slashed q}\gamma^{\mu}\rangle,
\qquad
q_{\mu}\langle\gamma^{\mu}{\slashed q}\gamma^{\nu}\rangle=
q_{\nu}\langle\gamma^{\mu}{\slashed q}\gamma^{\nu}\rangle=0.
\end{split}
\eeq

\noindent
As a result only the tensor structures that are odd in $k$ and in $q$ and are antisymmetric with respect to transposition of indices  $(\alpha,\beta) \leftrightarrow(\beta,\alpha)$ and $(\mu,\nu)\leftrightarrow(\nu,\mu)$ give contributions to HFS. All tensor structures containing $k_{\alpha}$, $k_{\beta}$, $q_{\mu}$, and $q_{\nu}$ do not contribute to HFS. Then for our purposes the LBL scattering tensor depends only on two independent structures

\beq  \label{struc-1a}
\begin{split}
{\cal M}^{(1)}_{\alpha \beta \mu \nu}(k,q)&=
\frac{k\cdot q}{2}(g_{\mu\alpha}g_{\nu\beta}- g_{\mu\beta}g_{\nu\alpha}),
\\
{\cal M}^{(2)}_{\alpha \beta \mu \nu}(k,q)& =
\frac{1}{4}(g_{\mu\alpha}k_{\nu}q_{\beta} - g_{\mu\beta}\,k_{\nu}q_{\alpha}
+ g_{\nu\beta}k_{\mu}q_{\alpha}- g_{\nu\alpha}k_{\mu}q_{\beta}).
\end{split}
\eeq

\noindent
Due to symmetries of the electron and muon factors in \eq{skobki} all terms in the tensor structures on the RHS  in \eq{struc-1a} give identical contributions to HFS, and in the calculations below we will use more compact expressions

\beq  \label{struc-1b}
{\cal M}^{(1)}_{\alpha \beta \mu \nu}(k,q)=(k\cdot q)g_{\mu\alpha}g_{\nu\beta},
\qquad
{\cal M}^{(2)}_{\alpha \beta \mu \nu}(k,q)=g_{\mu\alpha}k_{\nu}q_{\beta}.
\eeq

\noindent
The general expression for the LBL scattering tensor in \eq{lblintegral} can be obtained by representing the ladder and crossed traces in \eq{laddtr} and \eq{crstr} in the form

\beq  \label{otdeli-denom}
L_{\alpha \beta \mu \nu}=\frac{{\tilde L}_{\alpha \beta \mu\nu}}{D_1^2D_2D_3},
\qquad
C_{\alpha \beta \mu \nu} =\frac{{\tilde C}_{\alpha \beta \mu \nu}}{D_1D_2D_3D_4},
\eeq

\noindent
where

\beq \label{denom}
D_1 = p^2-m^2,
\quad
D_2=(p-q)^2-m^2,
\quad
D_3=(p-k)^2-m^2,
\quad
D_4=(p-q-k)^2-m^2.
\eeq

\noindent
Combining denominators in the expression for the crossed diagram with four different denominators in \eq{denom} with the help of the Feynman parameters $x, y, z,$ we obtain

\beq  \label{unden}
(1-x)D_4 + x\biggl\{(1-y)D_3 + y\Bigl[zD_2 + (1-z)D_1\Bigr]\biggr\}= (p-K)^2 -\Omega(x,y,z,\xi),
\eeq

\noindent
where

\beq
\label{bukva-Omega}
\begin{split}
&K=k(1-xy) + q (1-x+xyz),
\\
&\Omega(x,y,z,\xi)
=m^2-k^2xy(1-xy)- q^2x(1-yz)(1-x+xyz)
\\&
- 2(k\cdot q)xy[1-x-z(1-xy)]\xi.
\end{split}
\eeq

\noindent
The additional parameter $\xi$ is equal one in \eq{unden} and is introduced here for further convenience, we will explain its role later.  There is no denominator $D_4$ in the expression for the ladder diagram in \eq{otdeli-denom}  but the expressions in \eq{unden} and \eq{bukva-Omega} at $x=1$ are still suitable for calculations with the ladder diagram. To obtain an explicit expression for the LBL scattering tensor in \eq{lblintegral} it remains to calculate the integral over the loop momentum and to write the result in terms of the independent tensor structures in \eq{struc-1b}. The final expression for $S_{\alpha\beta\mu\nu}$ is rather cumbersome and will be presented below when we will use it for calculation of the single-logarithmic contribution in \eq{structure}. We do not need that general expression for discussion and calculation of the logarithm squared terms below.

\subsection{The Region of Strongly Ordered Momenta}

Large logarithmic contributions to HFS arise from logarithmic integrations in the region of strongly ordered momenta $m\ll k\ll q\ll M$ where the expression in \eq{lblskel} simplifies

\beq \label{logint}
\Delta E\simeq\frac{\alpha^2(Z\alpha)}{\pi^3}\frac{m}{M}E_F  \left(-\frac{3}{64}\right)
\int \frac{{d^4k}}{i\pi^2k^4}\frac{\langle\gamma^{\alpha}{\slashed k}\gamma^{\beta}\rangle}{k^2}
\int \frac{{d^4q}}{i\pi^2q^4}
\frac{M^2q^2}{q^4-4M^2q_0^2}\langle\gamma^{\mu}{\slashed q}\gamma^{\nu}\rangle
S_{\alpha \beta \mu \nu}.
\eeq

\noindent
The factor with the large mass $M$ is of order one when $q\ll M$ and integrations over $q$ and $k$ are logarithmic only if the LBL scattering tensor supplies a factor $k/q$. There are many ways how such factor arises in the expression for $S_{\alpha \beta \mu \nu}$. The leading contribution of this type could arise from the logarithmic integration over the loop momentum $p$ in the region

\beq \label{uslovie-0}
m\ll k\ll p\ll q\ll M.
\eeq

\noindent
The integrand in the expression for the crossed diagram in \eq{crstr} contains large $q^2$ in the denominator in the region of strongly ordered loop momenta  what makes integration over $q$ nonlogarithmic. Only the ladder diagrams in Fig. \ref{lblrec} could generate logarithm cubed contribution in this region. Expanding  over $q$ the ladder contribution to the LBL scattering tensor $S_{\alpha \beta \mu \nu}$  in \eq{lblintegral} we obtain

\beq \label{L-log3}
\begin{split}
{\cal L}_{\alpha \beta \mu \nu}
&=\int \frac{d^4p}{i\pi^2}L^{\alpha\beta\mu\nu}
=\int \frac{{d^4p}}{i\pi^2}Tr \left(\gamma_{\mu}\frac{1}{{\slashed p}-{\slashed q}-m}
\gamma_{\nu}\frac{1}{{\slashed p}-m}\gamma_{\beta}\frac{1}{{\slashed p}-{\slashed k}-m}
\gamma_{\alpha}\frac{1}{{\slashed p}-m}\right)
\\
&\simeq-\int \frac{{d^4p}}{i\pi^2}\frac{1}{q^2 p^8}
 Tr\left(\gamma_{\mu}{\slashed q}\gamma_{\nu}{\slashed p}\gamma_{\beta}{\slashed p}{\slashed k}{\slashed p}\gamma_{\alpha}{\slashed p}\right).
\end{split}
\eeq

\noindent
Projection of the last trace on the structures in \eq{struc-1a} is zero after averaging over directions of vector $p_\mu$. Hence, leading term in the expansion  of the LBL tensor does not contain $(k/q)\ln(q/k)$, and the logarithm cubed contribution to HFS does not arise despite the logarithmic nature of all integrations in the region of strongly ordered loop momenta in \eq{uslovie-0}.

\subsection{Logarithm Squared Contribution}

The logarithm squared contributions arise in two integration regions

\beq \label{uslovie-less}
m\ll k\sim p\leq\sigma\leq q\ll M,\quad
m\ll k\leq\sigma\leq p\sim q\ll M,
\eeq

\noindent
that are obtained from the region of strongly ordered loop momenta in \eq{uslovie-0}, when we lift the strong ordering requirement and allow two of the three loop momenta to be of the same order. For calculational purposes we introduce an auxiliary parameter $m\ll \sigma\ll M$ that separates the regions of large and small momenta and should cancel in the final results. As we have seen the LBL scattering tensor does not generate leading logarithmic terms of the type  $(k/q)\ln(q/k)$, and hence logarithms $(k/q)\ln(\sigma/k)$ and $(k/q)\ln(q/\sigma)$ do not arise in the large and small integration momenta regions in \eq{uslovie-less}.

To isolate the logarithm squared contributions we expand ladder and crossed contributions to the LBL scattering tensor in \eq{lblintegral} in the regions of small and large momenta in \eq{uslovie-less} and look for contributions of order $k/q$. The ladder contributions of this type arise from expansions both in the small and large momentum regions

\beq  \label{lad-less}
{\cal L}_{\alpha \beta \mu \nu}[p\le \sigma]\simeq \int \frac{{d^4p}}{i\pi^2}Tr \left(\gamma_{\mu}\frac{1}{-{\slashed q}}\gamma_{\nu}\frac{1}{{\slashed p}}\gamma_{\beta}\frac{1}{{\slashed p}-{\slashed k}}\gamma_{\alpha}\frac{1}{{\slashed p}}\right),
\eeq
\beq  \label{lad-more}
{\cal L}_{\alpha \beta \mu \nu}[p \ge \sigma] \simeq \int \frac{{d^4p}}{i\pi^2}Tr\left(\gamma_{\mu}\frac{1}{{\slashed p}-{\slashed q}}
\gamma_{\nu}\frac{1}{{\slashed p}}\gamma_{\beta}\frac{1}{{\slashed p}}{\slashed k} \frac{1}{{\slashed p}}\gamma_{\alpha}\frac{1}{{\slashed p}}\right).
\eeq

\noindent
The low momenta integral in \eq{lad-less} is superficially linearly divergent at the upper limit. This is an artificial divergence introduced in the integral when have thrown away momentum $p$ in comparison with $q$ in one of the denominators. We will deal with this divergence below extracting contributions of order $k/q$  from the integral in \eq{lad-less}. Notice also that in the large momenta region the integral in \eq{lad-more} generates a contribution of the form $k/q$ after expansion of the integrand up to the second order in $k/p$.

The contribution of the small integration momenta region for the crossed diagram in \eq{crstr} is suppressed as $\sigma^2/q^2$ due to large $q^2$ in the denominator of the crossed diagram in \eq{crstr}. The total leading logarithmic contribution to HFS connected with this diagram arises only from the large momenta region, where the terms of order $k/q$ arise, like in \eq{lad-more}, after expansion of the integrand up to the second order in $k/p$

\beq  \label{cross2}
{\cal C}_{\alpha \beta \mu \nu}\simeq \int \frac{{d^4p}}{i\pi^2}Tr \left(\gamma_{\mu}\frac{1}{{\slashed p}-{\slashed q}}
\gamma_{\beta}\frac{1}{{\slashed p}-{\slashed q}}\gamma_{\nu}
\frac{1}{{\slashed p}}{\slashed k}\frac{1}{{\slashed p}}\gamma_{\alpha}\frac{1}{{\slashed p}}
+
\gamma_{\mu}\frac{1}{{\slashed p}-{\slashed q}}\gamma_{\beta}\frac{1}{{\slashed p}-{\slashed q}}{\slashed k}\frac{1}{{\slashed p}-{\slashed q}}\gamma_{\nu}\frac{1}{{\slashed p}}\gamma_{\alpha}\frac{1}{{\slashed p}}\right).
\eeq

Now we are ready to calculate the logarithm squared contributions to HFS. We rationalize and combine denominators in \eq{lad-less}-\eq{cross2} like in \eq{denom}-\eq{bukva-Omega}, and extract the contributions proportional to $k/q$ from the respective integrals. Let us illustrate necessary transformations considering as an  example calculation of the small momenta ladder contribution in \eq{lad-less}. In this case we need only one Feynman variable $y$ in \eq{unden}, other variables are fixed, $x=1$, $z=0$. We obtain

\beq \label{boxless-2-1}
{\cal L}_{\alpha \beta \mu \nu}[p\le \sigma]
\simeq\frac{1}{q^2}\int\limits_{|p|\le\sigma}\frac{{d^4p}}{i\pi^2}
\int_0^1 {dy}\frac{2y}{\left[\left(p-k(1-y)\right)^2+k^2y(1-y)\right]^3}
Tr\left[\gamma_{\mu}{\slashed q}\gamma_{\nu}{\slashed p}\gamma_{\beta}({\slashed p}-{\slashed k})\gamma_{\alpha}{\slashed p}\right].
\eeq

\noindent
Next follows shift of the integration momentum $p\to p+k(1-y)$ in this formally linearly ultraviolet divergent integral. As usual a finite surface term arises after such shift (see, e.g., \cite{blp}). Further transformations are pretty standard, we calculate the trace, make the Wick rotation, preserve only the contributions proportional to $k/q$, calculate the integrals, extract the terms proportional to the tensor structures in \eq{struc-1b}, and obtain a finite result

\beq \label{boxless-tot2}
{\cal L}_{\alpha \beta \mu \nu}[p\le \sigma]=
\frac{4}{q^2}\left[(k\cdot q)g_{\mu \alpha}g_{\beta \nu}
- 2g_{\mu \alpha}q_{\beta}k_{\nu}\,\right].
\eeq

\noindent
Notice that the coefficient before the formally linearly divergent contribution turned into zero automatically. Calculation of the contributions in \eq{lad-more} and \eq{cross2} is no more difficult, and we obtain

\beq  \label{boxmore-tot}
{\cal L}_{\alpha \beta \mu \nu}[p \ge \sigma]=\frac{4}{q^2}g_{\mu\alpha}k_{\nu}q_{\beta},
\eeq

\beq  \label{ckross-tot}
{\cal C}_{\alpha \beta \mu \nu} =\frac{8}{q^2}
g_{\mu\alpha}\left[(k\cdot q)\,g_{\nu\beta}- 3k_{\nu}q_{\beta}\right].
\eeq

Finally, the total contribution to the LBL scattering tensor $S_{\alpha \beta \mu \nu}$ of the type $k/q$ arising in the region in \eq{uslovie-less} is the sum of the contributions in \eq{boxless-tot2}-\eq{ckross-tot}

\beq \label{lbltotal}
S_{\alpha \beta \mu \nu}=
2{\cal L}_{\alpha \beta \mu \nu}[p \le \sigma]
+2{\cal L}_{\alpha \beta \mu \nu}[p \ge \sigma]+{\cal C}_{\alpha \beta \mu \nu}
\simeq \frac{16}{q^2} \left[(k\cdot q)g_{\mu \alpha}g_{\beta \nu}
- 2g_{\mu \alpha}q_{\beta}k_{\nu}\right].
\eeq

\noindent
Comparing the small momenta contribution to the LBL tensor $2{\cal L}_{\alpha \beta \mu \nu}[p \le \sigma]$ (see \eq{boxless-tot2}) with the final expression above we see that the small momenta contribution to $S_{\alpha \beta \mu \nu}$ is equal the large momenta contribution. Next we substitute the LBL tensor in \eq{lbltotal} in the expression for HFS in \eq{logint} and  contract Lorentz indices

\beq  \label{ugolki-1}
\langle\gamma^{\alpha}{\slashed k}\gamma^{\beta}\rangle\langle\gamma^{\mu}{\slashed q}\gamma^{\nu}\rangle
g_{\mu \alpha}\, g_{\nu \beta}= -\frac{8}{3}\left[~2(k\cdot q)+ k_0q_0 \right],
\eeq

\beq  \label{ugolki-2}
\langle\gamma^{\alpha}{\slashed k}\gamma^{\beta}\rangle\langle\gamma^{\mu}{\slashed q}\gamma^{\nu}\rangle g_{\mu \alpha}k_{\nu}q_{\beta} = \frac{4}{3}\left[k^2q^2-(k\cdot q)^2 + k_0^2q^2
+ k^2q_0^2-2(k\cdot q)k_0q_0 \right].
\eeq

\noindent
Averaging over directions of $k$ we obtain the ladder and crossed contributions to the LBL tensor

\beq \label{lad-prosto}
\langle\gamma^{\alpha}{\slashed k}\gamma^{\beta}\rangle\langle\gamma^{\mu}{\slashed q}\gamma^{\nu}\rangle
{\cal L}_{\alpha \beta \mu \nu} =\langle\gamma^{\alpha}{\slashed k}\gamma^{\beta}\rangle\langle\gamma^{\mu}{\slashed q}\gamma^{\nu}\rangle
\frac{4}{q^2}\left[(k\cdot q)g_{\mu\alpha}g_{\nu\beta}- g_{\mu\alpha}k_{\nu}q_{\beta}\right]
=-\frac{16}{3} k^2 \frac{2q^2+q_0^2}{q^2},
\eeq

\beq \label{cro-prosto}
\langle\gamma^{\alpha}{\slashed k}\gamma^{\beta}\rangle\langle\gamma^{\mu}{\slashed q}\gamma^{\nu}\rangle
{\cal C}_{\alpha \beta \mu \nu} =
\langle\gamma^{\alpha}{\slashed k}\gamma^{\beta}\rangle\langle\gamma^{\mu}{\slashed q}\gamma^{\nu}\rangle
\frac{8}{q^2} g_{\mu\alpha}\left[(k\cdot q)g_{\nu\beta} - 3 k_{\nu}q_{\beta} \right]
=-4 \frac{16}{3} k^2 \frac{2q^2+q_0^2}{q^2}.
\eeq

\noindent
Then the total LBL scattering tensor in this regime has the form

\beq \label{sumlc}
\langle\gamma^{\alpha}{\slashed k}\gamma^{\beta}\rangle\langle\gamma^{\mu}{\slashed q}\gamma^{\nu}\rangle
(2{\cal L}_{\alpha \beta \mu \nu}+{\cal C}_{\alpha \beta \mu \nu})
=-32 k^2 \frac{2q^2+q_0^2}{q^2},
\eeq

\noindent
and the logarithm squared contribution to HFS can be written as

\beq
\Delta E =\frac{\alpha^2(Z\alpha)}{\pi^3}\frac{m}{M}E_F \frac{3}{2}
\int \frac{{d^4q}}{i\pi^2q^4}\frac{M^2(2q^2+q_0^2)}{q^4-4M^2q_0^2}\int \frac{{d^4k}}{i\pi^2k^4}.
\eeq

The logarithm squared contribution arises in the integration region $m^2\ll k^2\ll q^2\ll M^2$, and we calculate it using, after the Wick rotation, four-dimensional spherical coordinates  and preserving only the logarithmic contribution

\beq \label{log2again}
\begin{split}
\Delta E&=\frac{\alpha^2(Z\alpha)}{\pi^3}\frac{m}{M}E_F \left(-\frac{3}{2}\right)
\frac{2}{\pi}\int_{0}^{\pi}{d\theta}\sin^2{\theta}
\int_{m^2}^{M^2} \frac{{dq^2}}{q^2}\frac{M^2(2+\cos^2{\theta})}{q^2+4M^2\cos^2{\theta}}
\int_{m^2}^{q^2} \frac{{dk^2}}{k^2}
\\
&
\simeq\frac{9}{4}\frac{\alpha^2(Z\alpha)}{\pi^3}\frac{m}{M}E_F\ln^2{\frac{M}{m}}.
\end{split}
\eeq

\noindent
This logarithm squared contribution was obtained in \cite{eks89}.

\begin{figure}[htb]
\includegraphics[height=1.cm]{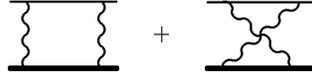}
\caption{\label{twoph}Diagrams with two-photon exchanges}
\end{figure}

Another way \cite{kse1990} to calculate this contribution is to notice that it is intimately connected with the two-loop renormalization of the axial vector current first calculated by Adler long time ago \cite{adler}. Consider the leading recoil correction to HFS generated by the graphs with two-photon exchanges in Fig. \ref{twoph}. Respective contribution to HFS is given by the expression in \eq{lblskel} without the LBL scattering block. The antisymmetric electron-line and and muon-line spin factors in \eq{skobki} can be written in the form

\beq  \label{struc-1}
\langle\gamma^{\alpha}\gamma^\delta\gamma^{\beta}\rangle_{(e)} \longrightarrow
i\epsilon^{\rho \alpha \delta \beta}\langle\gamma_{\rho}\gamma_5 \rangle_{(e)},
\qquad
\langle\gamma^{\mu}\gamma^\sigma\gamma^{\nu}\rangle_{(\mu)} \longrightarrow i\epsilon^{\lambda \mu \sigma \nu}\langle\gamma_{\lambda}\gamma_5\rangle_{(\mu)}.
\eeq

\noindent
Then the leading recoil contribution to HFS has the form

\beq \label{leadingre}
\Delta E =\frac{3}{4}\frac{Z\alpha}{\pi}\frac{m}{M}E_F \langle\gamma^{\lambda}\gamma_5\rangle_{(\mu)}\int_m^M \frac{{dq}}{q}\langle\gamma_{\lambda}\gamma_5 \rangle_{(e)}.
\eeq

\begin{figure}[htb]
\includegraphics[height=2.cm]{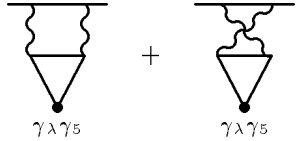}
\caption{\label{axial}
Two-loop axial current renormalization}
\end{figure}

\noindent
We see that the leading recoil contribution to HFS is determined by the matrix element of the electron axial current calculated at the characteristic virtuality $q$. The first radiative correction to this matrix element is of order $\alpha^2\ln (q^2/m^2)$ and arises at two loops \cite{adler} (see Fig. \ref{axial})

\beq \label{axren}
j_\lambda^5\to j_\lambda^5\left[1-\frac{3}{4}\left(\frac{\alpha}{\pi}\right)^2\ln\frac{q^2}{m^2}\right].
\eeq

\noindent
Substituting this renormalization factor in the expression for the leading recoil correction in \eq{leadingre} we obtain the leading recoil correction accompanied by the logarithm squared contribution in \eq{log2again}

\beq
\Delta E=-\frac{3}{4}\frac{Z\alpha}{\pi}\frac{m}{M}E_F \ln\frac{M}{m}
+\frac{9}{4}\frac{\alpha^2(Z\alpha)}{\pi^3}\frac{m}{M}E_F\ln^2{\frac{M}{m}}.
\eeq

\noindent
We see that the logarithm squared contribution is an observable effect of the axial current renormalization.

Let us clarify the connection between the calculation of this contribution based on the consideration of the LBL scattering tensor and the approach with the axial current renormalization. The diagrams in Fig. \ref{axial} are naively linearly divergent and the result in \eq{axren} implies a gauge invariant regularization. On the other hand contribution of the diagrams in Fig. \ref{lblrec} to HFS is ultraviolet finite. Hence, for the purpose of calculation of the logarithm squared contribution replacement of the two triangle diagrams in Fig. \ref{axial}  by the three box diagrams in Fig. \ref{lblrec} can be considered as a gauge invariant regularization. Let us see how this regularization works. Consider first the integration region where $k\sim p\leq\sigma\leq q$. In this region the lower fermion line in the LBL scattering box in the first two diagrams in Fig. \ref{lblrec} effectively shrinks to a point,  and these diagrams turn into the axial current diagrams in Fig. \ref{axial}. We have already calculated contribution to the LBL scattering tensor $2{\cal L}_{\alpha \beta \mu \nu}[p\le \sigma]$ generated in this region, see \eq{boxless-tot2}. This contribution is exactly one half of the total contribution to the LBL scattering tensor in \eq{lbltotal}. The other half (see \eq{boxmore-tot} and \eq{ckross-tot}) arises in the integration region $k\leq \sigma\leq p\sim q$, where both lower loops in all diagrams in Fig. \ref{lblrec} effectively shrink to a point. Therefore, contribution to the axial current renormalization generated in this region can be considered as a pure regularization effect, for example contribution of a heavy regularizing fermion  that survives when the fermion mass goes to infinity.

\section{Single-Logarithmic Contribution}

Calculation of the single-logarithmic contribution to HFS requires more accurate treatment of the LBL scattering tensor in \eq{lblintegral}. Calculating traces in \eq{otdeli-denom} we obtain

\beq
\begin{split}
\tilde L^{\alpha \beta \mu \nu}& = 8D_1^2 g^{\mu \alpha}g^{\nu \beta}
+ 16D_1 g^{\mu \alpha} \left[p^{\nu}q^{\beta} + k^{\nu}p^{\beta}
+ k^{\nu} q^{\beta}-p^{\nu}p^{\beta}\right]
- 8D_1 g^{\mu \alpha}g^{\nu \beta}
\\
&\times \left[(p\cdot q) + (p\cdot k) + (k\cdot q)\right]+ 32 g^{\mu \alpha}\left[(k\cdot q) p^{\nu}p^{\beta}
- (p\cdot q) k^{\nu}p^{\beta}
-(p\cdot k) p^{\nu} q^{\beta}\right]
\\
&+ 16 (p\cdot k)(p\cdot q) g^{\mu \alpha}g^{\nu \beta}
- 32 p^{\mu} p^{\alpha} k^{\nu} q^{\beta},
\end{split}
\eeq

\noindent
and

\beq
\begin{split}
\tilde C^{\alpha \beta \mu \nu}
&=8D_1g^{\mu \alpha}\left[-3k^{\nu} q^{\beta} + k^{\nu} p^{\beta} + p^{\nu} q^{\beta}\right]
- 8D_2  g^{\mu \alpha}k^{\nu}p^{\beta}
- 8D_3  g^{\mu \alpha}p^{\nu}q^{\beta}
\\
&+ 16(k\cdot q)g^{\mu \alpha}p^{\nu}p^{\beta}
- 16(k\cdot q)g^{\mu \alpha}k^{\nu}p^{\beta}
- 16(k\cdot q)g^{\mu \alpha}p^{\nu}q^{\beta}
+ 16p\cdot(k+q)g^{\mu \alpha}k^{\nu}q^{\beta}.
\end{split}
\eeq

\noindent
Then after the shift in \eq{unden} and calculation of the loop integral the LBL scattering tensor can be written in the form

\beq \label{gauginvblb}
S^{\alpha\beta\mu\nu}=2{\cal L}^{\alpha\beta\mu\nu}+{\cal C}^{\alpha\beta\mu\nu},
\eeq

\noindent
where (see \eq{otdeli-denom})

\beq \label{lblladd}
\begin{split}
{\cal L}^{\alpha\beta\mu\nu}&=\int \frac{d^4p}{\pi^2i}L^{\alpha\beta\mu\nu}
=-16\int_0^1 {dy} \int_0^1 {dz}\frac{y^2(1-y)z}{\Omega(1,y,z,1)}
[(k\cdot q)g_{\mu\alpha}g_{\nu\beta}
- g_{\mu\alpha}k_{\nu}q_{\beta}]
\\
&+16\int_0^1 {dy} \int_0^1 {dz}\biggl\{\frac{y(1-2y) +2y^2z}{\Omega(1,y,z,1)}
\\
&
+y^2(1-z)\frac{k^2(1-y)^2 + q^2y^2z^2}{\Omega^2(1,y,z,1)}\biggr\}
\biggl[(k\cdot q)g_{\mu\alpha}g_{\nu\beta}
- 2g_{\mu\alpha}k_{\nu}q_{\beta}\biggr]
\\
&+16\int_0^1 {dy} \int_0^1 {dz}\Biggl\{\biggl[\frac{m^2y}{\Omega(1,y,z,1)}
-\int_0^1 {d\xi}\frac{2(k\cdot q)^2 y^3(1-y)^2z^2\xi}
{\Omega^2(1,y,z,\xi)}
\\
&+y^3(1-y)z(1-z)\frac{k^2q^2 + (k\cdot q)^2}{\Omega^2(1,y,z,1)}\biggr] g_{\mu\alpha}g_{\nu\beta}
\\
&
-y^3(1-y)z(1-z)\frac{2(k\cdot q)}{\Omega^2(1,y,z,1)}g_{\mu\alpha}k_{\nu}q_{\beta}\Biggr\},
\end{split}
\eeq

\noindent
and

\beq \label{lblcross}
\begin{split}
{\cal C}^{\alpha\beta\mu\nu}
&=\int\frac{d^4p}{\pi^2i}C^{\alpha\beta\mu\nu}
=8\int_0^1dx\int_0^1 {dy} \int_0^1 {dz}x^2y\Biggl\{
-\frac{(k\cdot q)g_{\mu \alpha}g_{\nu \beta}-3g_{\mu \alpha}k_{\nu}q_{\beta}}{\Omega(x,y,z,1)}
\\
&
+\frac{4m^2 - 2k\cdot q xy[1-x-z(1-xy)]}{\Omega^2(x,y,z,1)}
g_{\mu \alpha}k_{\nu}q_{\beta}\Biggr\}.
\end{split}
\eeq

\noindent
The additional parameter $\xi$ in \eq{lblladd} arises when we separate the ultraviolet divergence in the logarithmically divergent integral

\beq
\label{div-trans}
\begin{split}
&\int \frac{{d^4p}}{i\pi^2}\frac{p^2}{[p^2-\Omega(1,y,z,1)]^3}
=\int\frac{{d^4p}}{i\pi^2}\frac{p^2}{[p^2-\Omega(1,y,z,0)]^3}
\\
&+\int \frac{{d^4p}}{i\pi^2}p^2\biggl[\frac{1}{[p^2-\Omega(1,y,z,1)]^3}
-\frac{1}{(p^2-\Omega(1,y,z,0))^3}\biggr]
\\
&=\biggl[\ln{\frac{\Lambda^2}{\Omega(1,y,z,0)}}-\frac32\biggr]
-\int_0^1 {d\xi}\frac{2(k\cdot q)y(1-y)z}{\Omega(1,y,z,\xi)}.
\end{split}
\eeq

\noindent
As was explained above the LBL scattering tensor is contacted with the odd in $k$ and in $q$ tensor structures and the even ultraviolet divergent term in the square brackets does not contribute to HFS and can be thrown away.

It is easy to check that the terms on the RHS in the first lines in \eq{lblladd} and \eq{lblcross} reproduce the leading ladder and crossed contributions to the LBL scattering tensor in \eq{lbltotal} and generate the logarithm squared contribution to HFS in \eq{log2again}.

At the next step we calculate the integral  over momentum $k$ of the upper photons in \eq{lblskel}

\beq  \label{T-def1}
\begin{split}
T(q^2,q_0)&=\frac12\int \frac{{d^4k}}{i\pi^2k^4}
\biggl(\frac{1}{k^2+2mk_0}+\frac{1}{k^2-2mk_0}\biggl) \langle\gamma^{\alpha}{\slashed k}\gamma^{\beta}\rangle
\langle\gamma^{\mu}{\slashed q}\gamma^{\nu}\rangle
S_{\alpha \beta \mu \nu}
\\
&=\langle\gamma^{\mu}{\slashed q}\gamma^{\nu}\rangle\int \frac{{d^4k}}{i\pi^2k^4}
\frac{ \langle\gamma^{\alpha}{\slashed k}\gamma^{\beta}\rangle}{k^2-2mk_0}S_{\alpha \beta \mu \nu},
\end{split}
\eeq

\noindent
where we used the symmetry of the integrand under simultaneous substitution $k\to-k$ and $q\to-q$ to get rid of the second term in the first brackets on the RHS in the first line.  In terms of $T(q^2,q_0)$ the contribution to HFS in \eq{lblskel} has the form

\beq  \label{CROSS-1}
\Delta E =\frac{\alpha^2(Z\alpha)}{\pi^3}\frac{m}{M}E_F\biggl(-\frac{3M^2}{128}\biggr)
\int \frac{{d^4q}}{i\pi^2q^4}\biggl(\frac{1}{q^2+2Mq_0}+\frac{1}{q^2-2Mq_0}\biggl) T(q^2,q_0).
\eeq

We calculate $T(q^2,q_0)$ combining denominators with the help of additional Feynman parameters $u$ and $t$ (the four-vector $Q$ and the scalar $\Delta$ on the RHS depend on the parameters $x,y,z,\xi,u$, and $t$)

\beq  \label{denom-ut}
(1-u)\Bigl[(1-t)k^2 + t(k^2-2mk_0)\Bigr]+u\Biggl[\frac{\Omega(x,y,z,\xi)}{-xy(1-xy)}\Biggr]
=(k-Q)^2 - \Delta,
\eeq

\noindent
where

\beq  \label{Delta}
\begin{split}
&\Delta = g\Bigl[-q^2 + 2bq_0 + a^2\Bigr],
\quad
a^2=\frac{1}{g}\biggl[\tau^2+\frac{m^2u}{xy(1-xy)}\biggr],\quad
b=\frac{\tau d}{g},
\\
&d=\xi u\biggl[z- \frac{1-x}{1-xy}\biggr],\qquad \tau =m(1-u)t,
\qquad
g=g_0-d^2,
\\
&g_0=\frac{u(1-yz)(1-x+xyz)}{y(1-xy)}.
\end{split}
\eeq

\noindent
The four-vector $Q$ has the form

\beq  \label{vector-Q}
Q_{\mu} =d q_{\mu} + \tau_{\mu},
\eeq

\noindent
where $\tau_{\mu} =(\tau,{\bm 0})$ and to calculate the integral over $k$ we shift the integration variable

\beq  \label{sdvizhka-natural}
k_{\mu}\longrightarrow k_{\mu} + Q_{\mu} =k_{\mu} +dq_{\mu} +\tau_{\mu}.
\eeq

\noindent
Contractions of matrix structures in \eq{T-def1} are calculated using relationships in \eq{ugolki-1} and \eq{ugolki-2}. Five different tensor structures arise in calculations, and after the shift of integration variable in \eq{sdvizhka-natural} they reduce to

\beq  \label{shiftstruc-1}
\begin{split}
&\langle\gamma^{\alpha}{\slashed k}\gamma^{\beta}\rangle\langle\gamma^{\mu}{\slashed q}\gamma^{\nu}\rangle
 g_{\mu \alpha} g_{\nu \beta}
\longrightarrow -\frac83 (2q^2+q_0^2)-8q_0 \tau,
\\
&\langle\gamma^{\alpha}{\slashed k}\gamma^{\beta}\rangle\langle\gamma^{\mu}{\slashed q}\gamma^{\nu}\rangle
 (k\cdot q)g_{\mu \alpha} g_{\nu \beta}
\longrightarrow -\frac83\biggl(\frac14 k^2 + q^2d^2\biggr)  (2q^2+q_0^2)
-8q_0^2 \tau^2 -\frac83(5q^2+q_0^2) d  q_0\tau ,
\\
&\langle\gamma^{\alpha}{\slashed k}\gamma^{\beta}\rangle\langle\gamma^{\mu}{\slashed q}\gamma^{\nu}\rangle
g_{\mu \alpha} k_{\nu}q_{\beta}
\longrightarrow \frac{2}{3} k^2 (2q^2+q_0^2)
+\frac{8}{3} (q^2 - q_0^2) \tau^2 ,
\\
&\langle\gamma^{\alpha}{\slashed k}\gamma^{\beta}\rangle\langle\gamma^{\mu}{\slashed q}\gamma^{\nu}\rangle
 (k\cdot q)^2g_{\mu \alpha} g_{\nu \beta}
\longrightarrow-\frac{8}{3}\biggl(\frac34 k^2 + q^2d^2\biggr)
(2q^2+q_0^2) q^2 d
-\frac{8}{3}(8q^2 + q_0^2)\tau^2  q_0^2 d
\\
&-\frac{8}{3}\biggl(\frac14 k^2 + q^2d^2\biggr) (7q^2+2q_0^2) q_0\tau
-8 q_0^3  \tau^3,
\\
&\langle\gamma^{\alpha}{\slashed k}\gamma^{\beta}\rangle\langle\gamma^{\mu}{\slashed q}\gamma^{\nu}\rangle
 (k\cdot q) g_{\mu \alpha} k_{\nu}q_{\beta}
\longrightarrow \biggl[\frac{2}{3} k^2 (2q^2+q_0^2)
+\frac{8}{3}\tau^2 (q^2 - q_0^2)\biggr] \Bigl(q^2d +q_0\tau \Bigr) .
\end{split}
\eeq

Now we are ready to obtain explicit integral representations for the ladder $T_{L}(q^2, q_0)$ and crossed $T_{C}(q^2, q_0)$ diagram contributions to the function $T(q^2,q_0)$

\beq
T(q^2,q_0)=2T_{L}(q^2,q_0)+T_{C}(q^2,q_0).
\eeq

\noindent
The ladder contribution can be written as a sum of nine integrals

\beq  \label{genint-L}
T_{L}(q^2, q_0)=\frac{128}{3}\int_0^1 {dy} \int_0^1 {dz}\int_0^1 {du}\int_0^1 {dt}
\sum_i {\cal T}_{L,i}(y, z, u, t, q^2, q_0),
\eeq

\noindent
where ($x=1$ in all formulae below and $\xi=1$ in all expressions except ${\cal T}_{L,6}$)

\beq \label{calTL-1}
\begin{split}
{\cal T}_{L,1}& =  yz(1-t)(1-u)^2
\Biggr\{\biggl[\frac{1}{\Delta}-\frac{q^2 d^2}{\Delta^2} \biggr] (2q^2+q_0^2)
- \frac{ (q^2+2q_0^2)\tau^2}{\Delta^2}
- \frac{ q_0  (5q^2+q_0^2) \tau d}{\Delta^2} \Biggr\},
\end{split}
\eeq
\beq \label{calTL-2}
\begin{split}
{\cal T}_{L,2}& =\frac32 (2q^2+q_0^2)\Biggl\{-\frac{(1-2y) +2yz}{1-y}
 \frac{(1-t)(1-u)^2}{\Delta}
\\
&+(1-z) \frac{u(1-u)}{\Delta}
-\frac{y^2z^2(1-z) q^2}{(1-y)^2}
\frac{(1-t) u(1-u)^2}{\Delta^2}\Biggr\},
\end{split}
\eeq
\beq \label{calTL-3}
\begin{split}
{\cal T}_{L,3}& =\Biggl\{\frac{(1-2y) +2yz}{1-y} \frac{(1-t)(1-u)^2}{\Delta^2}
-(1-z) \frac{u(1-u)}{\Delta^2}
\\
&+2 \frac{y^2z^2(1-z) q^2}{(1-y)^2}
\frac{(1-t) u(1-u)^2}{\Delta^3}\Biggr\} (2q^2+q_0^2) q^2d^2,
\end{split}
\eeq
\beq \label{calTL-4}
\begin{split}
{\cal T}_{L,4}& =
\Biggl\{\frac{(1-2y) +2yz}{1-y} \frac{(1-t)(1-u)^2}{\Delta^2}
\\
&-(1-z) \frac{u(1-u)}{\Delta^2}
+2 \frac{y^2z^2(1-z) q^2}{(1-y)^2}
\frac{(1-t) u(1-u)^2}{\Delta^3}\Biggr\}
\\
&\times\biggl[(2q^2+q_0^2)  \tau^2 + q_0 (5q^2+q_0^2) \tau  d\biggr],
\end{split}
\eeq
\beq \label{calTL-5}
\begin{split}
{\cal T}_{L,5}& =\frac{m^2}{1-y}\frac{(1-t)(1-u)^2}{\Delta^2}
\Bigl[(2q^2+q_0^2) d + 3 q_0\tau \Bigr],
\end{split}
\eeq
\beq \label{calTL-6}
\begin{split}
{\cal T}_{L,6}&  =4\int_0^1 {d\xi}\xi  yz^2 (1-t) u(1-u)^2
\\
&\times\Biggl\{\biggl[\frac34 \frac{1}{\Delta_{\xi}^2} - \frac{q^2d^2_{\xi}}{\Delta_{\xi}^3}\biggr]
(2q^2+q_0^2) q^2 d_{\xi}
- \frac{\tau^2  q_0^2 d_{\xi}}{\Delta_{\xi}^3}  (8q^2+q_0^2)
\\
&+\biggl[\frac14 \frac{1}{\Delta_{\xi}^2} - \frac{q^2d^2_{\xi}}{\Delta_{\xi}^3}\biggr]
(7q^2+2q_0^2) q_0\tau - \frac{3 q_0^3 \tau^3}{\Delta_{\xi}^3}\Biggr\},
\end{split}
\eeq
\beq \label{calTL-7}
\begin{split}
{\cal T}_{L,7}& =-\frac{yz(1-z)}{1-y}\frac{q^2 u(1-u)}{\Delta^2}
\Bigl[(2q^2+q_0^2) d + 3 q_0\tau \Bigr],
\end{split}
\eeq
\beq \label{calTL-8}
\begin{split}
{\cal T}_{L,8}& = 2 \frac{yz(1-z)}{1-y} (1-t) u(1-u)^2
\Biggl\{\biggl[-\frac34 \frac{1}{\Delta^2} + \frac{q^2d^2}{\Delta^3}\biggr]
(2q^2+q_0^2) q^2 d
\\
&+ \frac{\tau^2  q_0^2 d}{\Delta^3}  (8q^2+q_0^2)
+\biggl[-\frac14 \frac{1}{\Delta^2} + \frac{q^2d^2}{\Delta^3}\biggr]
(7q^2+2q_0^2) q_0\tau + \frac{3 q_0^3 \tau^3}{\Delta^3}\Biggr\},
\end{split}
\eeq
\beq
\label{calTL-9}
\begin{split}
{\cal T}_{L,9}&  =4 \frac{yz(1-z)}{1-y}(1-t) u(1-u)^2
\biggl[-\frac{1}{4} \frac{1}{\Delta^2} (2q^2+q_0^2) q^2d
\\
&+q^2(q^2-q_0^2) \frac{\tau^2 d}{\Delta^3}
- \frac{1}{4} \frac{1}{\Delta^2}(2q^2+q_0^2) q_0 \tau
+q_0(q^2-q_0^2) \frac{\tau^3}{\Delta^3}\biggr].
\end{split}
\eeq

\noindent
The crossed diagram contribution can be written as a sum of three integrals

\beq  \label{genint-C}
T_{C}(q^2, q_0)=\frac{128}{3}\int_0^1 {dx}\int_0^1 {dy} \int_0^1 {dz}\int_0^1 {du}\int_0^1 {dt}
\sum_i {\cal T}_{C,i}(x, y, z, u, t, q^2, q_0),
\eeq

\noindent
where ($\xi=1$ in all formulae below)

\beq \label{calTC-1}
{\cal T}_{C,1} =\frac12 \frac{x(1-t)(1-u)^2}{1-xy}
\Biggr[(2q^2+q_0^2) \biggl[\frac{2}{\Delta}-\frac{q^2 d^2}{\Delta^2}\biggr]
- 3\frac{ q^2\tau^2}{\Delta^2}
- \frac{ q_0  (5q^2+q_0^2) \tau d}{\Delta^2} \Biggr],
\eeq
\beq  \label{calTC-2}
{\cal T}_{C,2} = \frac{x(1-t)(1-u)^2}{1-xy}
\frac{u m^2}{xy(1-xy)}
\biggl[\frac{2q^2+q_0^2}{\Delta^2}
-4 \frac{(q^2-q_0^2) \tau^2}{\Delta^3} \biggr],
\eeq
\beq
\label{calTC-3}
\begin{split}
{\cal T}_{C,3}& =\frac12 \frac{x(1-t)(1-u)^2}{1-xy}
 \Biggl[(2q^2+q_0^2) \frac{q^2 d^2}{\Delta^2}
-4(q^2-q_0^2) \frac{ q^2 \tau^2  d^2}{\Delta^3}
\\
&+(2q^2+q_0^2) \frac{q_0 \tau d}{\Delta^2}
-4(q^2-q_0^2) \frac{q_0 \tau^3  d}{\Delta^3}\Biggl].
\end{split}
\eeq

All logarithmic contributions to HFS can be obtained from the large momentum expansion of $T(q^2,q_0)$. The leading term in this expansion we already obtained in \eq{sumlc}

\beq \label{T-leadasymp}
T=-32\int \frac{{d^4k}}{i\pi^2k^4}\frac{q^2+2q_0^2}{q^2}
\simeq -32\int_{m^2}^{-q^2} \frac{{dk^2}}{k^2}~\frac{q^2+2q_0^2}{q^2}
\simeq-32\frac{q^2+2q_0^2}{q^2}\ln{\frac{-q^2}{m^2}}.
\eeq

\noindent
We are going to calculate the next terms in the large $q$ expansion of $T(q^2,q_0)$

\beq \label{T-log-const}
T= -32\frac{q^2+2q_0^2}{q^2}\ln{\frac{-q^2}{m^2}}
+\kappa_1\frac{2q^2+q_0^2}{q^2}+ \kappa_2\frac{q^2+2q_0^2}{q^2},
\eeq

\noindent
where $\kappa_1$ and $\kappa_2$ are the numerical coefficients to be calculated. The numerators of the last two terms should contain any two linear independent combinations of $q^2$ and $q_0^2$. We have chosen $2q^2+q_0^2$ and $q^2+2q_0^2$ because the first one contains  the same combination of momenta that arises in the leading term of the expansion in \eq{T-leadasymp}, and the second structure does not generate logarithm of the mass ratio after integration over $q$.

\begin{table}[htb]
\caption{Leading terms in the ladder diagram expansion}
\begin{ruledtabular}
\begin{tabular}{clc}
${\cal T}_{L,i}$ &  $-\frac{3}{16}\kappa_{1i}$ & $-\frac{3}{16}\kappa_{2i}$     \\
\colrule
${\cal T}_{L, 1}$       &~~$\ln{\frac{-q^2}{m^2}} + 8\zeta{(3)} - 8$ & $1$          \\
${\cal T}_{L, 2a}$    &    $-6\ln{\frac{-q^2}{m^2}} - 9$ & $0$   \\
${\cal T}_{L, 2b}$    &    $~~3\ln{\frac{-q^2}{m^2}} -24\zeta{(3)}- \frac{4\pi^2}{3} +35$  & $0$     \\
${\cal T}_{L, 2c}$    &   $~~3\ln{\frac{-q^2}{m^2}} -24\zeta{(3)} +33$ & $0$    \\
${\cal T}_{L, 3}$    &     $~~96\zeta{(3)}- 116$ & $0$   \\
${\cal T}_{L, 4}$    &    $~~\frac{4\pi^2}{9} - \frac{8}{3}$ & $0$ \\
${\cal T}_{L, 5}$    &    $~~0$   & $0$  \\
${\cal T}_{L, 6}$    &   $-16\zeta{(3)} + 19$    & $0$     \\
${\cal T}_{L, 7}$    &   $-16\zeta{(3)} + 20$  & $0$    \\
${\cal T}_{L, 8}$    &   $-40\zeta{(3)} + \frac{97}{2}$ & $0$   \\
${\cal T}_{L, 9}$    &   $~~16\zeta{(3)} - 19$   & $0$   \\
\colrule
${\cal T}_{L}$ &   $\ln{\frac{-q^2}{m^2}} - \frac{8\pi^2}{9}+ \frac{5}{6}$ & $1$
\end{tabular}
\end{ruledtabular}
\label{ladderasymp}
\end{table}

The integral representations in \eq{calTL-1}-\eq{calTL-9}, and in \eq{calTC-1}-\eq{calTC-3} are rather cumbersome and we simplify them before integration over the Feynman parameters. It turns out that in calculations of the asymptotic expansion in \eq{T-log-const} we can omit the term $2bq_0$ in the denominator $\Delta$

\beq  \label{Delta-simple}
\Delta=g(q^2 +2bq_0 + a^2)\to g(q^2 + a^2)
\equiv \widetilde{\Delta}.
\eeq

\noindent
To justify this simplification it is sufficient to notice that the term $2bq_0$ in the denominator $\Delta$ arises when we combine the subleading term $2mk_0$ from the electron propagator with the subleading term $2k\cdot q$ from the denominator $\Omega$ in \eq{denom-ut}. The denominator $\widetilde{\Delta}$ depends only on the Lorentz invariant momentum squared $q^2$ what makes the calculations easier. Even after this simplification the integrals over the Feynman remain unwieldy, especially in the case of the crossed diagrams when they contain an extra Feynman parameter $x$. They become more manageable if we notice that the remnant of the scalar product $k\cdot q$ in the denominator  $\Omega$ survives not only as the term $2bq_0$ in  \eq{Delta-simple} but also as the second term in the factor $g$ (see definitions in \eq{Delta}, below we show explicitly only dependence of $g$, $d$, and $\widetilde\Delta$ on one parameter $\xi$)

\beq  \label{g-total}
g(\xi) = g(0) - d^2
= \frac{u(1-yz)(1-x+xyz)}{y(1-xy)} - \xi^2u^2\biggl[ z- \frac{1-x}{1-xy}\biggr]^2.
\eeq

\noindent
The leading logarithmic term in \eq{T-log-const} arises when we simply omit this second term, $g(\xi)\to g(0)$. To calculate the subleading terms in the asymptotic expansion in \eq{T-log-const} we represent all terms with the denominator $\widetilde\Delta(\xi=1)$ in \eq{Delta-simple} in the form

\beq \label{simplifrden}
\frac{1}{\widetilde\Delta(\xi=1)}=\frac{1}{\widetilde\Delta(\xi=0)}
+\left[\frac{1}{\widetilde\Delta(\xi=1)}
-\frac{1}{\widetilde\Delta(\xi=0)}\right]=
\frac{1}{\widetilde\Delta(\xi=0)}-\int_0^1d\xi\frac{2\xi q^2d^2(\xi)}{\widetilde\Delta(\xi)}.
\eeq

\noindent
It is easier to calculate separately the integrals with the two terms on the RHS than the integral with the denominator $\widetilde\Delta$ on the LHS side.  Unlike the expansions in \eq{boxmore-tot}-\eq{lbltotal} the integrals with the denominator $\widetilde\Delta$ substituted  by the terms on the RHS  are well suited for calculation of the subleading terms in the asymptotic expansions in the kinematics described in \eq{uslovie-less}. After tedious calculations  we obtained analytic expressions for the subleading terms in the asymptotic expansion of the integrals in \eq{calTL-1}-\eq{calTL-9} and in \eq{calTC-1}-\eq{calTC-3}. These terms are collected in Tables \ref{ladderasymp} and \ref{crossasymp}, and the asymptotic expansions for the ladder and crossed diagrams have the form

\beq  \label{ladder-asymp}
T_{L}= -\frac{16}{3}\frac{2q^2+q_0^2}{q^2}
\biggl[\ln{\frac{-q^2}{m^2}} - \frac{8\pi^2}{9}+ \frac{5}{6}\biggr]
-\frac{16}{3}\frac{q^2+2q_0^2}{q^2},
\eeq

\medskip

\beq  \label{cross-asymp}
T_{C} = -\frac{64}{3} \frac{2q^2+q_0^2}{q^2}
\biggl[ \ln{\frac{-q^2}{m^2}} -2\zeta{(3)} + \frac{8}{3} \biggr]
- \frac{32}{3}\frac{q^2+2q_0^2}{q^2}.
\eeq

\noindent
Then the total ultraviolet asymptotic expansion in \eq{T-log-const} acquires the form

\beq \label{total-asymp}
T = -32\frac{2q^2+q_0^2}{q^2}
\biggl[\ln{\frac{-q^2}{m^2}}-\frac43\zeta{(3)} - \frac{8\pi^2}{27}+ \frac{37}{18}\biggr]-\frac{64}{3}\frac{q^2+2q_0^2}{q^2}.
\eeq

Our next task is to calculate the single-logarithmic contributions generated by the first two terms in the expansion in \eq{total-asymp}.  We substitute the asymptotic expansions in \eq{ladder-asymp} and \eq{cross-asymp} in the expression for HFS in \eq{CROSS-1}. One can prove that the last terms with the numerator $q^2+2q_0^2$ do not generate logarithmic contributions and all double- and single-logarithmic contributions are generated by the terms proportional $2q^2+q_0^2$. The leading logarithmic term in \eq{total-asymp} generates not only the logarithm squared contribution in \eq{log2again} but also additional single-logarithmic terms. Analytic calculation of single-logarithmic terms is performed  with the help of auxiliary integration formulae collected in \cite{egs1998}. We obtain ladder and crossed diagram logarithmic contributions to HFS in the form

\beq  \label{L-finasymp}
2\Delta E_{L} \simeq 2 \frac{\alpha^2(Z\alpha)}{\pi^3}\frac{m}{M}E_F
\biggl[\frac{3}{8} \ln^2{\frac{{M}}{m}}
+ \biggl(- \frac{\pi^2}{3} + \frac{23}{16}\biggr)\ln{\frac{{M}}{m}}\biggr],
\eeq

\beq  \label{C-finasymp}
\Delta E_{C}\simeq \frac{\alpha^2(Z\alpha)}{\pi^3}\frac{m}{M}E_F  \biggl[\frac{3}{8}\ln^2{\frac{{M}}{m}}
+ \biggl(- 3\zeta{(3)} + \frac{17}{2}\biggr)\ln{\frac{{M}}{m}}\biggr].
\eeq

\begin{table}[htb]
\caption{Leading terms in the crossed diagram expansion}
\begin{ruledtabular}
\begin{tabular}{clc}
$ {\cal T}_{C, i}$ & $-\frac{3}{64}\kappa_{1i}$ & $-\frac{3}{64}\kappa_{2i}$
\\
\colrule
& &\\[-2.2ex]
${\cal T}_{C, 1}$    &     $~~\ln{\frac{-q^2}{m^2}}  - \frac{\pi^2}{3}+ \frac72$ & $\frac12$         \\
${\cal T}_{C, 2}$    &    $- \frac72\zeta{(3)} + \frac{\pi^2}{3}+1$ & $0$ \\
${\cal T}_{C, 3}$    &     $~~ \frac32\zeta{(3)} - \frac{11}{6}$ & $0$ \\
\colrule
${\cal T}_{C}$&   $~~\ln{\frac{-q^2}{m^2}} - 2\zeta{(3)} + \frac{8}{3}$ &  $\frac12$
\end{tabular}
\end{ruledtabular}
\label{crossasymp}
\end{table}

And finally the total logarithmic radiative-recoil contribution to HFS generated by the gauge invariant set of of three-loop diagrams with the LBL insertions has the form

\beq  \label{finasymp}
\Delta E = \frac{\alpha^2(Z\alpha)}{\pi^3}\frac{m}{M}E_F \biggl[\frac{9}{4} \ln^2{\frac{{M}}{m}}
+ \biggl(-3\zeta{(3)}- \frac{2\pi^2}{3} + \frac{91}{8}\biggr) \ln{\frac{{M}}{m}}\biggr].
\eeq

\section{Conclusions}

Other single-logarithmic radiative-recoil contributions to HFS were calculated earlier  \cite{egs01,egs03,egs04,esprd2009,esprl2009,esjetp2010}

\beq
\Delta E=\frac{\alpha^3}{\pi^3}\frac{m}{M}E_F\left[3\zeta(3)-6\pi^2\ln2+\pi^2-8\right]\ln\frac{M}{m}.
\eeq

\noindent
Combining this contribution with the result obtained above in \eq{finasymp} we obtain the total result for all known three-loop radiative-recoil single-logarithmic corrections to HFS

\beq
\Delta E_{tot}=\frac{\alpha^3}{\pi^3}\frac{m}{M}E_F
\left(-6\pi^2\ln2+\frac{\pi^2}{3}+\frac{27}{8}\right)\ln\frac{M}{m}.
\eeq

\noindent
As was explained in the Introduction, the current goal of the HFS theory in muonium is to reduce the theoretical uncertainty below 10 Hz. The result above is  a step in this direction. Work on calculation of the remaining three-loop single-logarithmic  and nonlogarithmic contributions to HFS is now is progress.

\begin{acknowledgments}

This work was supported by the NSF grant PHY-1066054. The work of V. S. was also supported in part by the RFBR grant 12-02-00313 and by the DFG grant GZ: HA 1457/7-2.

\end{acknowledgments}

\end{document}